\documentclass{aa}

\usepackage{graphicx}
\usepackage[varg]{txfonts}

\begin{document}

   \title{Bayesian mass and age estimates for transiting exoplanet host
stars\thanks{The source code and stellar model grids for our method  are
available at the CDS via anonymous ftp to cdsarc.u-strasbg.fr
(130.79.128.5) or via http://cdsweb.u-strasbg.fr/cgi-bin/qcat?J/A+A/.}}


   \author{P.~F.~L.~Maxted\inst{1}
          \and
          A.~M.~Serenelli\inst{2}
          \and
          J.~Southworth\inst{1}
          }

   \institute{Astrophysics Group,  Keele University, Keele, Staffordshire ST5
5BG, UK\\
              \email{p.maxted@keele.ac.uk}
         \and
Instituto de Ciencias del Espacio (CSIC-IEEC), Facultad de Ciencias,
Campus UAB, 08193, Bellaterra, Spain
             }

   \date{Received ; accepted }

 
  \abstract
   {The mean density of a star transited by a planet, brown dwarf or low mass
star can be accurately measured from its light curve. This measurement can be
combined with other observations to estimate its mass and age by
comparison with stellar models.}
   {Our aim is to calculate the posterior probability distributions for the
mass and age of a star given its density, effective temperature, metallicity and
luminosity.}
   {We computed a large grid of stellar models that densely sample the
appropriate mass and metallicity range. The posterior probability distributions
are calculated using a Markov-chain Monte-Carlo method. The method has been
validated by comparison to the results of other stellar models and by applying
the method to stars in eclipsing binary systems with accurately measured
masses and radii. We have explored the sensitivity of our results to the
assumed values of the mixing-length parameter, $\alpha_\mathrm{MLT}$, and
initial helium mass fraction, $Y$.}
   {For a star with a mass of 0.9\,M$_{\sun}$ and an age of 4\,Gyr our method
recovers the mass of the star with a precision of 2\% and the age to within
25\% based on the density, effective temperature and metallicity predicted by a
range of different stellar models. The masses of stars in eclipsing binaries
are recovered to within the calculated uncertainties (typically 5\%) in about
90\% of cases. There is a tendency for the masses to be underestimated by
about 0.1\,M$_{\sun}$ for some stars with rotation periods $P_\mathrm{rot} <
7$~d. }
{Our method makes it straightforward to determine accurately the joint
posterior probability distribution for the mass and age of a  star eclipsed by
a planet or other dark body based on its observed properties and a
state-of-the art set of stellar models. }
   \keywords{stars: solar-type -- binaries: eclipsing -- planetary systems
               }
   \maketitle
%

\section{Introduction}

 Studies of extrasolar planets rely on a good understanding of the stars that
they orbit. To estimate the mass and radius of an extrasolar planet that
transits its host star we require  an estimate for the mass of the star. The
mass of the star will also strongly influence the planet's environment, e.g.,
the spectrum and intensity of the stellar flux intercepted by the planet and
the nature and strength of the tidal interaction between the star and the
planet. The ages of planet host stars are used to investigate the lifetimes of
planets and the time scales for  tidal interactions between the planet and the
star \citep[e.g.,][]{2010ApJ...725.1995M, 2010A+A...512A..77L,
2011MNRAS.415..605B}.

 The analysis of the light curve produced by a planetary transit yields an
accurate estimate for the radius of the star relative to the semi-major
axis of the planet's orbit, $R_{\star}/a$, provided that the eccentricity of
the orbit is known. This estimate can be combined with Kepler's laws to
estimate the density of the host star. The density can be combined with
estimates for the effective temperature and metallicity of the star to infer a
mass and age for the star by comparison with stellar models or an empirical
calibration of stellar mass. In general, the comparison with stellar models is
done using a maximum likelihood method, i.e., taking the mass and age of the
evolution track and isochrone that give the best fit to the observed density
(or $R_{\star}/a$) and effective temperature, estimated either by a
least-squares fit to the observed properties or ``by-eye''.  

 Maximum-likelihood estimates can be strongly biased in cases where the
mapping between the observed parameters  and the parameters of interest is
non-linear. This is certainly the case for stellar ages because the observed
parameters change very little during the main-sequence phase, but there are
large changes to the observed properties during the rapid evolution of a star
away from the main sequence. This can produce a ``terminal age bias'', where
the ``best-fit'' method applied to a sample of stars produces a distribution of
ages that is {\it a priori} very unlikely, i.e., too few main-sequence stars
and many stars in regions of the model paramater space corresponding to
short-lived evolutionary phases. This problem is particularly acute for cases
where the uncertainties on the observed parameters are large compared to the
change in observed properties during the main sequence phase. This is
often the case for the age estimates of single stars based mainly on surface
gravity measured from the stellar spectrum or the absolute magnitude derived
from the measured parallax and stellar flux. Bayesian methods that account for
the {\it a priori} distribution of stellar ages have been developed to deal
with this problem in spectroscopic stellar surveys 
\citep{2005A+A...436..127J,2004MNRAS.351..487P,
2013MNRAS.429.3645S, 2014arXiv1408.3409S}.
\citet{2009A+A...502..695P} applied a similar Bayesian approach in their
study of the HD~80606 planetary system, but there have been few other examples
of this approach in exoplanet studies. This may be because there is currently
no software available to the exoplanet research community that can be used to
apply these Bayesian methods and that is straightforward to use.

 In general, the terminal age bias is expected to be less severe for planet
host stars than for single stars because stellar density measurements based on
the analysis of a planetary transit are usually more precise than surface
gravity estimates based on the analysis of a stellar spectrum or the
luminosity derived from parallax and flux measurements. The mean stellar
density is also more sensitive to the change in radius of a star as it evolves
away from the main sequence. However, precise stellar densities can also cause
problems because the broad sampling in mass, age and metallicity used for many
grids of stellar models can produce poor sampling of the observed parameter
spacing, i.e., the typical difference in stellar density between adjacent
model grid points can be much larger than the uncertainty on the observed
value. This can produce systematic errors due to interpolation, particularly
for stars near the end of their main-sequence evolution where the evolution
tracks have a complex behaviour that is very sensitive to age, mass and
composition.  This also makes it difficult to make reliable estimates of the
uncertainties on the mass and age. One method sometimes employed to estimate
the uncertainties is to look for the mass and age range of all the models that
pass within the estimated errors on the observed values, e.g., all the models
within the 1-$\sigma$ error bars. This approach can be misleading because the
errors on the mass and age are often strongly non-Gaussian and highly
correlated . It also possible to miss some combinations of mass, age and
composition that provide a reasonable match to the observed properties of the
star but that are not sampled by the stellar model grid or that fall just
outside the 1-$\sigma$ error bars, particularly when the fitting is done
by-eye. 

 The recently-launched European Space Agency mission GAIA
\citep{2001A+A...369..339P} will measure parallaxes and optical fluxes for
many stars that are transited by planets, brown-dwarfs and low-mass stars. The
luminosity measurement derived from these observations can be an additional
useful constraint on the mass and age of the star. If the density, effective
temperature and composition of the star are also known then finding the
best-fit mass and age of the star becomes an over-determined problem for
stellar models with fixed helium abundance and mixing-length parameter, i.e.,
there are more observables than unknowns. If the best-fit to the observed
parameters is poor then this opens up the possibility of using these stars to
explore whether the assumed values of the helium abundance and mixing-length
parameter, or some other factor,  can be adjusted to improve the agreement
between the stellar models and real stars.

 To deal with these issues we have developed a Markov-chain Monte Carlo (MCMC)
method that calculates the posterior probability distribution for the mass and
age of a star from its observed mean density and other observable quantities
using  a grid of stellar models that densely samples the relevant parameter
space. We have validated our method by applying it to data derived from
different stellar models and by applying it to stars in eclipsing binary stars
with precisely measured masses and radii. We have also quantified the
systematic error in the estimated mass and age due to the variations in the
assumed helium abundance and convective mixing length parameter. The method
has been implemented as a program called {\sc bagemass} that we have made
available for general use. 

\section{Method} 
\subsection{Stellar models and grid interpolation}

 Our method uses a grid of models for single stars  produced with the
{\sc garstec} stellar evolution code \citep{2008Ap&SS.316...99W}. The methods
used to calculate the stellar model grid are described in
\citet{2013MNRAS.429.3645S} so we only summarise the main features of the
models and some differences to that description here. 

 {\sc garstec} uses the  FreeEOS\footnote{\url{http://freeeos.sourceforge.net}}
equation of state \citep{2003ApJ...588..862C} and standard mixing length
theory for convection \citep{1990sse..book.....K}. The mixing length
parameter used to calculate the model grid is $\alpha_\mathrm{MLT}=1.78$. With
this value of $\alpha_\mathrm{MLT}$ {\sc garstec}  reproduces the observed
properties of the present day Sun assuming that the composition is that given
by \citet{1998SSRv...85..161G}, the overall initial solar metallicity is
$Z_{\sun} = 0.01826$, and the initial solar helium abundance is
$Y_{\sun}=0.26646$. These are slightly different to the value in
\citet{2013MNRAS.429.3645S} because we have included additional mixing
below the convective zone in order reduce the effect of gravitational settling
and so to better match the properties of metal-poor stars. Due to the effects
of microscopic diffusion, the initial solar composition corresponds to an
initial iron abundance  $[\mathrm{Fe/H}]_{\mathrm{i}} = +0.06$.

 The stellar model grid covers the mass range 0.6\,\mbox{M$_{\sun}$}\ to
2.0\,\mbox{M$_{\sun}$}\ in steps of 0.02\,\mbox{M$_{\sun}$}. The grid of
initial metallicity values covers the range $\mathrm{[Fe/H]}_{\mathrm{i}} =
-0.75$ to $-0.05$ in steps of 0.1\,dex and the range
$\mathrm{[Fe/H]}_{\mathrm{i}} = -0.05$ to $+0.55$ in 0.05\,dex steps.  The
initial composition of the models is computed assuming a cosmic
helium-to-metal enrichment $\Delta Y/ \Delta Z =
(Y_{\sun}-Y_\mathrm{BBN})/Z_{\sun}$, where $Y_\mathrm{BBN}=0.2485$ is the
primordial helium abundance due to big-bang nucleosynthesis
\citep{2010JCAP...04..029S}, so $\Delta Y/ \Delta Z = 0.984$.
  
 The initial abundance of all elements is scaled according to the value of
$\mathrm{[Fe/H]}_{\mathrm{i}}$. For each value of initial mass and
$\mathrm{[Fe/H]}_{\mathrm{i}}$ we  extracted the output from {\sc garstec} at
999 ages from the end of the pre-main-sequence phase up to an age of 17.5\,Gyr
or a maximum radius of 3\,\mbox{R$_{\sun}$}, whichever occurs first. We define
the zero-age main sequence (ZAMS) to be the time at which the star reaches its
minimum luminosity and measure all ages relative to this time.

 To obtain the properties of a star from our  model grid at arbitrary mass,
$\mathrm{[Fe/H]}_{\mathrm{i}}$ and age we use the {\sc pspline} implementation
of the cubic spline interpolation
algorithm.\footnote{\url{http://w3.pppl.gov/ntcc/PSPLINE}}. We interpolated
the stellar evolution tracks for stars that reach the terminal-age main
sequence (TAMS) on to two grids, one from the start of the evolution track to
the TAMS, and one that covers the post-main sequence evolution. The dividing
line between the two is set by the age at which the central hydrogen abundance
drops to 0. We use 999 grid points evenly distributed in age for each model
grid. For stars on the main sequence we interpolate between models as a
function of age in units of the main sequence lifetime at the specified values
of mass and $\mathrm{[Fe/H]}_{\mathrm{i}}$. The interpolating variable for the
post-main sequence properties is the age since the TAMS measured in units of
the total time covered by the model grid. Splitting the grid of stellar models
in this way improves the accuracy of the interpolation near the terminal-age
main sequence.

 We used {\sc garstec} to calculate some models for solar metallicity at masses
half-way between those used for our method and then compared the interpolated
values to those calculated by {\sc garstec}. For the masses that we checked
(1.01\,\mbox{M$_{\sun}$} and 1.29\,\mbox{M$_{\sun}$}) we find that the maximum
error in the density is 0.01\,$\rho_{\sun}$, the maximum error in
\mbox{T$_\mathrm{eff}$} is 25\,K and the maximum error in $\log (L/L_{\sun})$
is 0.015. The worst agreement between the calculated and interpolated models
occurs at very young ages ($<0.01$\,Gyr) and, for the 1.29\,\mbox{M$_{\sun}$}
star, near the ``blue hook'' as the star evolves off the main sequence. Away
from these evoutionary phases the error in the interpolated values is at least
an order-of-magnitude smaller that these ``worst case'' values.

\subsection{Input data\label{MethodSection}}

 It is much easier to calculate the probability distribution functions of a
star's mass and age if the observed quantities can be assumed to be
independent. To enable us to make this assumption we define  the data to be a
vector of observed quantities \mbox{$\vec{d} = \left(\mbox{T$_\mathrm{eff}$}, \mbox{L$_{\star}$},
\mathrm{[Fe/H]}_{\mathrm{s}},\mbox{$\rho_{\star}$}\right)$}, where \mbox{T$_\mathrm{eff}$}\ is the effective
temperature of the star, \mbox{L$_{\star}$}\ its observed luminosity,
$\mathrm{[Fe/H]}_{\mathrm{s}}$ characterises the surface metal abundance and
\mbox{$\rho_{\star}$}\ is the stellar density.

 The density of stars that host a transiting extrasolar planet can be
determined directly from the analysis of the light curve if the eccentricity
is known. From Kepler's third law it follows that
\begin{equation*}
\mbox{$\rho_{\star}$} =
 \frac{3\mbox{M$_{\star}$}}{4\pi\mbox{R$_{\star}$}^3} = \frac{3\pi}{GP^2(1+q)}
 \left(\frac{a}{\mbox{R$_{\star}$}}\right)^3, 
\end{equation*}
where $P$ and $a$ are the period and semi-major axis of the Keplerian orbit
and $q = M_\mathrm{c}/\mbox{M$_{\star}$}$ is the mass ratio for a companion
with mass $M_\mathrm{c}$. The  quantity $(a/\mbox{R$_{\star}$})$ can be
determined with good precision from a high quality light curve alone if the
orbit is known to be circular since it depends only on the depth, width and
shape of the transit \citep{2003ApJ...585.1038S}. For non-circular orbits
the same argument applies but in this case the transit yields
$(d_t/\mbox{R$_{\star}$})$, where $d_t$ is the separation of the stars during
the transit, so the eccentricity and the longitude of periastron must be
measured from the spectroscopic orbit or some other method so that the ratio
$d_t/a$ can be determined. The error in the mass ratio will give a negligible
contribution to the uncertainty on \mbox{$\rho_{\star}$} for any star where
the presence of an extrasolar planet has been confirmed by radial velocity
observations.  The same argument applies to brown-dwarf or low mass stellar
companions, but some care is needed to make an accurate estimate of the mass
ratio, e.g., via the mass function using an initial estimate of the
stellar mass, and the additional uncertainty in \mbox{$\rho_{\star}$} should
be accounted for.

 Most of the stars currently known to host exoplanets do not have accurately
measured trigonometrical parallaxes. For these stars we set
$\log(\mbox{L$_{\star}$}/\mbox{L$_{\sun}$}) = 0\pm 5$ so that this parameter
has a negligible influence on the results. For stars with measured parallaxes
the observed luminosity is $\mbox{L$_{\star}$} =  4\pi d^2 f_{\oplus}$ where
$f_{\oplus}$, the flux from the star at the top of the Earth's atmosphere
corrected for reddening, and $d$, the distance to the star measured from its
trigonometrical parallax. The values of \mbox{T$_\mathrm{eff}$}\ and
$\mathrm{[Fe/H]}_{\mathrm{s}}$ can be derived from the analysis of a high
quality  stellar spectrum. These  may be less accurate than the quoted
precision because of the approximations used in the stellar model atmospheres
and the uncertainties in atomic data used in the analysis. For that reason we
set lower limits of 80\,K on the standard error for \mbox{T$_\mathrm{eff}$}\
and 0.07 dex for the standard error on $\mathrm{[Fe/H]}_{\mathrm{s}}$
\citep{2010MNRAS.405.1907B}.  For stars similar to the Sun a change of
80\,K in the assumed value of \mbox{T$_\mathrm{eff}$}\ results in a change of
about 0.02 dex in the value of $\mathrm{[Fe/H]}_{\mathrm{s}}$ derived from the
analysis of the spectrum \citep{2013MNRAS.428.3164D}. This is at least a
factor of 3 lower than the minimum standard error that we have assumed for
$\mathrm{[Fe/H]}_{\mathrm{s}}$ so we ignore this weak correlation between
\mbox{T$_\mathrm{eff}$}\ and $\mathrm{[Fe/H]}_{\mathrm{s}}$. For a few stars
the effective temperature can be determined directly from $f_{\oplus}$ and the
directly measured angular diameter. The quoted uncertainties on these
directly-measured \mbox{T$_\mathrm{eff}$}\ values should not be used if they
are much smaller than about 80\,K because our method does not account for the
strong correlation between \mbox{T$_\mathrm{eff}$}\ and \mbox{L$_{\star}$} in
these cases and some allowance must be made for the uncertainties in the
effective temperature scale of the stellar models. 

\subsection{Calculation of the Bayesian mass and age estimates}
 The vector of model parameters that are used to predict the observed data is
\mbox{$\vec{m} = \left(\mbox{$\tau_{\star}$}, \mbox{M$_{\star}$},
\mathrm{[Fe/H]}_{\mathrm{i}}\right)$}, where $\mbox{$\tau_{\star}$}$,
$\mbox{M$_{\star}$}$ and $\mathrm{[Fe/H]}_{\mathrm{i}}$ are the age, mass and
initial metal abundance of the star, respectively.  The observed surface metal
abundance $\mathrm{[Fe/H]}_{\mathrm{s}}$  differs from the initial metal
abundance $\mathrm{[Fe/H]}_{\mathrm{i}}$ because of diffusion and mixing
processes in the star during its evolution. 

 We use the MCMC method to determine the probability distribution function
\mbox{$p(\vec{m}|\vec{d}) \propto {\cal L}(\vec{d}|\vec{m})p(\vec{m})$}. To
estimate the likelihood of observing the data $\vec{d}$ for a given model
$\vec{m}$ we use \mbox{${\cal L}(\vec{d}|\vec{m}) = \exp(-\chi^2/2)$}, where

\begin{eqnarray*}
\chi^2   = & \frac{\left(T_{\mathrm{eff}} - T_{\mathrm{eff,obs}}\right)^2}{\sigma_T^2} 
+ \frac{\left(\log(L_{\star}) -
\log(L_{\star,\mathrm{obs}})\right)^2}{\sigma_{\log L}^2}\\
 & + \frac{\left(\mathrm{[Fe/H]}_{\mathrm{s}} - \mathrm{[Fe/H]}_{\mathrm{s,obs}}\right)^2}
        {\sigma_{\mathrm{[Fe/H]}}^2}
  + \frac{\left(\rho_{\star} - \rho_{\star,\mathrm{obs}}\right)^2}{\sigma_{\rho}^2}.
\end{eqnarray*}
 In this expression for $\chi^2$ observed quantities are denoted by the
subscript `obs', their standard errors are $\sigma_T$, $\sigma_{\log L}$,
etc., and other quantities are derived from the model, as described above. In
cases where asymmetric error bars are quoted on values we use either the upper
or lower error bar, as appropriate, depending on whether the model value is
greater than or less than the observed value. The probability distribution
function \mbox{$p(\vec{m}) = p(\mbox{$\tau_{\star}$}) p(\mbox{M$_{\star}$})
p(\mathrm{[Fe/H]}_{\mathrm{i}})$} is the product of the individual priors on
each of the model parameters. The assumed prior on
$\mathrm{[Fe/H]}_{\mathrm{i}}$ normally has little effect because this
parameter is well constrained by the observed value of
$\mathrm{[Fe/H]}_{\mathrm{s}}$ so we generally use a `flat' prior on
$\mathrm{[Fe/H]}_{\mathrm{i}}$, i.e., a uniform distribution over the model
grid range. It is possible in our method to set a prior on M$_{\star}$ of the
form $p(M_{\star}) = e^{\alpha M_{\star}}$. The prior on \mbox{M$_{\star}$}\
is the product of the present day mass function for planet host stars, the
mass distribution of the target stars in the surveys that discovered these
planets and the sensitivity of these surveys as a function of stellar mass,
but since none of these factors  is well determined we normally use a
flat prior for \mbox{M$_{\star}$}, i.e, $\alpha=0$. We also use a flat prior
on \mbox{$\tau_{\star}$}\ over the range 0\,--\,17.5\,Gyr.   In general, and
for all the results below unless stated otherwise, we simply set set
$p(\vec{m}) = 1$ for all models where the parameters are within the range of
valid values. However, the implementation of our method does include the
option to set priors on [Fe/H] or $\tau_{\star}$ of the form \begin{equation*}
p(x) = \left\{ \begin{array}{ll}
e^{-(x-x_\mathrm{lo})^2/2\sigma_\mathrm{lo}^2} &,~ x< x_\mathrm{lo} \\
1 &,~ x_\mathrm{lo} < x < x_\mathrm{hi}\\
e^{-(x-x_\mathrm{hi})^2/2\sigma_\mathrm{hi}^2} &,~ x > x_\mathrm{hi}. \\
\end{array} \right. \end{equation*}

 We generate a set of points $\vec{m}_i$  (a {\it Markov chain}) with the
probability distribution \mbox{$p(\vec{m}|\vec{d})$} using a jump
probability distribution $f(\Delta\vec{m})$ that specifies how to generate a
trial point $\vec{m}^{\prime} = \vec{m}_i + \Delta\vec{m}$. The trial
point is always rejected if any of the model parameters are outside their
valid range. Otherwise, a point is always included in the chain if  ${\cal
L}(\vec{m}^{\prime}|\vec{d}) > {\cal L}(\vec{m}_i|\vec{d})$  and may
be included in the chain with probability ${\cal
L}(\vec{m}^{\prime}|\vec{d})/{\cal L}(\vec{m}_i|\vec{d})$ if ${\cal
L}(\vec{m}^{\prime}|\vec{d}) < {\cal L}(\vec{m}_i|\vec{d})$
(Metropolis-Hastings algorithm). If the trial point is accepted in the chain
then $\vec{m}_{i+1} = \vec{m}^{\prime}$, otherwise $\vec{m}_{i+1}
=\vec{m}_{i}$ \citep{2004PhRvD..69j3501T}. 

 We randomly sample 65,536 points uniformly distributed over the model grid
range and take the point with the lowest value of $\chi^2$ as the first point
in the chain. From this starting point, $\vec{m}_0$,  we find the step size
for each parameter such that $|\ln({\cal L}(\vec{m}_0|\vec{d}) - \ln({\cal
L}(\vec{m}_0^{(j)}|\vec{d})| \approx 0.5$, where $\vec{m}_0^{(j)}$ denotes a
set of model parameters that differs from $\vec{m}_0$ only in the value of one
parameter, $j$.  We then produce a Markov chain with 10,000 steps using a
multi-variate Gaussian distribution for $f(\Delta\vec{m})$ with the standard
deviation of each parameter set to this step size. This first Markov chain is
used to find an improved set of model parameters and the second half of this
chain is used to calculate the covariance matrix of the model parameters. We
then calculate the eigenvalues and eigenvectors of the covariance matrix,
i.e., the principal components of the chain. This enables us to determine a
set of transformed parameters, $\vec{q} = (q_1, q_2, q_3)$, that are
uncorrelated and where each of the $q_i$  have unit variance, as well as the
transformation from $\vec{q}$ to $\vec{m}$. We then produce a Markov chain
with 50,000 steps using a multi-variate Gaussian distribution for
$f(\Delta\vec{q})$ with unit standard deviation for each of the transformed
parameters. The first point of this second Markov chain is the set of model
parameters with the highest value of ${\cal L}(\vec{m}_i|\vec{d})$ from the
first Markov chain. This second Markov chain is the one used to estimate the
probability distribution function \mbox{$p(\vec{m}|\vec{d})$}. We use visual
inspection of the chains and the Gelman-Rubin  statistic \citep{GR92} to
ensure that the chains are well mixed, i.e., that they properly sample the
parameter space. The number of steps used in the two Markov chains can be
varied, e.g., longer chains can be used to ensure that the parameter space is
properly sampled in difficult cases. 

 Our algorithm is implemented as a {\sc fortran} program called {\sc bagemass}
that accompanies the on-line version of this article and that is also
available as an open source software
project.\footnote{\url{http://sourceforge.net/projects/bagemass}}

\section{Validation of the method}

\subsection{Comparison to other models}

 In Table~\ref{ModFitTable} we compare the predicted values of
\mbox{T$_\mathrm{eff}$}\ and \mbox{$\rho_{\star}$}\ from our {\sc garstec}
models for a star at an age of 4\,Gyr with an initial mass of 0.9\,M$_{\sun}$
and solar composition  to those of three other grids of stellar models. These
are all grids of ``standard stellar models' in the sense that they assume a
linear relation between helium enrichment and metallicity, and the mixing
length parameter is calibrated using a model of the Sun. The Dartmouth Stellar
Evolution Program (DSEP) model grid is described in
\citet{2008ApJS..178...89D}.  The ``VRSS 2006'' model grid is described in
\citet{2006ApJS..162..375V} and the Geneva 2012 models are described by
\citet{2012A+A...541A..41M}. The  VRSS 2006 models do not include diffusion
or gravitational settling of elements. 

 The range in \mbox{T$_\mathrm{eff}$}\ values is 65\,K, with the {\sc garstec}
models being at the top end of this range. The values of
\mbox{$\rho_{\star}$}\ vary by about 3.5\%, with the {\sc garstec} models
predicting the lowest density while the VRSS 2006 models predict the highest
density. The differences are mainly due to differences in the assumed solar
metallicity that is used for the zero-point of the [Fe/H] scale and the assumed
value of $\alpha_\mathrm{MLT}$ in each model grid.

 We used {\sc bagemass} to find the best-fitting (maximum-likelihood) mass and
age estimates for these model stars based on the values given in
Table~\ref{ModFitTable}. We assigned standard errors of 80\,K to
\mbox{T$_\mathrm{eff}$}\ and 0.07 dex to $\mathrm{[Fe/H]}_{\mathrm{i}}$, i.e.,
the minimum standard errors on these values that we use for the analysis of
real stars. We assumed that the error on \mbox{$\rho_{\star}$}\ is
$\pm0.001\rho_{\sun}$. The results are also shown in Table~\ref{ModFitTable}.
The results for the {\sc garstec} models show that our method is
self-consistent to better than the 1-per~cent level in recovering the stellar
mass and age of the star. Comparing our results to those of other models, we
see that the systematic error due to differences in the stellar models in the
recovered stellar mass is less than about 2\%. The systematic error in the
recovered age is larger ($\approx 25$\%) for this example. Note that these
figures may not apply to stars at different masses or ages, e.g., differences
in the treatment of convective overshooting can lead to large differences in
the predicted age for more massive stars.

\begin{table}
\caption{Maximum-likelihood mass and age estimates from {\sc bagemass} for
model stars with masses of 0.9\mbox{M$_{\sun}$}\,
$\mathrm{[Fe/H]}_{\mathrm{i}} = 0.0$ and an age of 4\,Gyr.}
\label{ModFitTable}
\centering
\begin{tabular}{@{}lrrrrr}
\hline\hline
Model &
  \multicolumn{1}{l}{\mbox{T$_\mathrm{eff}$}} &
  \multicolumn{1}{l}{$\mbox{$\rho_{\star}$}$} &
  \multicolumn{1}{l}{$\mathrm{[Fe/H]}_{\mathrm{s}}$} & 
  \multicolumn{1}{l}{Mass} &
  \multicolumn{1}{l}{Age}\\
  \multicolumn{1}{l}{} &
  \multicolumn{1}{c}{[K]} &
  \multicolumn{1}{c}{[$\rho_{\sun}$]} &
 &
  \multicolumn{1}{c}{[\mbox{M$_{\sun}$}]} &
  \multicolumn{1}{c}{[Gyr]}\\
\hline
{\sc garstec} & 5435 & 1.486 &$-0.035$ & 0.900 & 4.00 \\
DSEP 2008    & 5372 & 1.514 &$-0.044$ & 0.884 & 4.59 \\
VRSS 2006    & 5388 & 1.539 & 0  & 0.903 & 3.15 \\
Geneva 2012  & 5370 & 1.517 & $-0.019$ & 0.891 & 4.18 \\
\hline
 \end{tabular}   
 \end{table}

\subsection{Eclipsing binary stars}

\begin{figure}
\resizebox{\hsize}{!}{\includegraphics{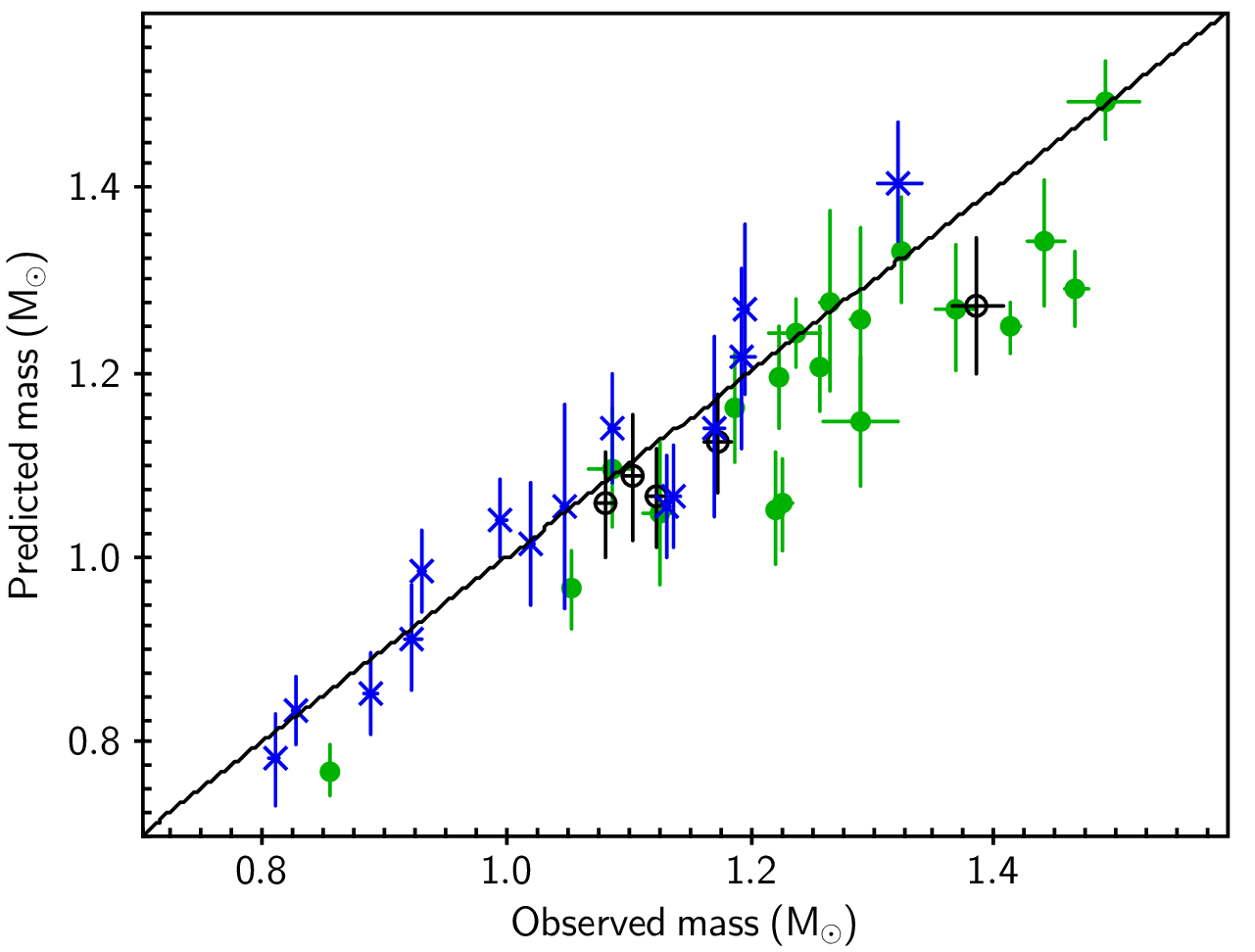}}
\caption{Bayesian mass estimates for stars in detached eclipsing binary star systems.
Symbols/colours denote the following orbital period ranges:
$P_\mathrm{orb} < 6$\,d -- filled circles/green; $6 < P < 12$\,d -- open
circles/black;  $P>12$\,d -- crosses/blue.
\label{debmassFig}}
\end{figure}

\begin{figure}
\resizebox{\hsize}{!}{\includegraphics{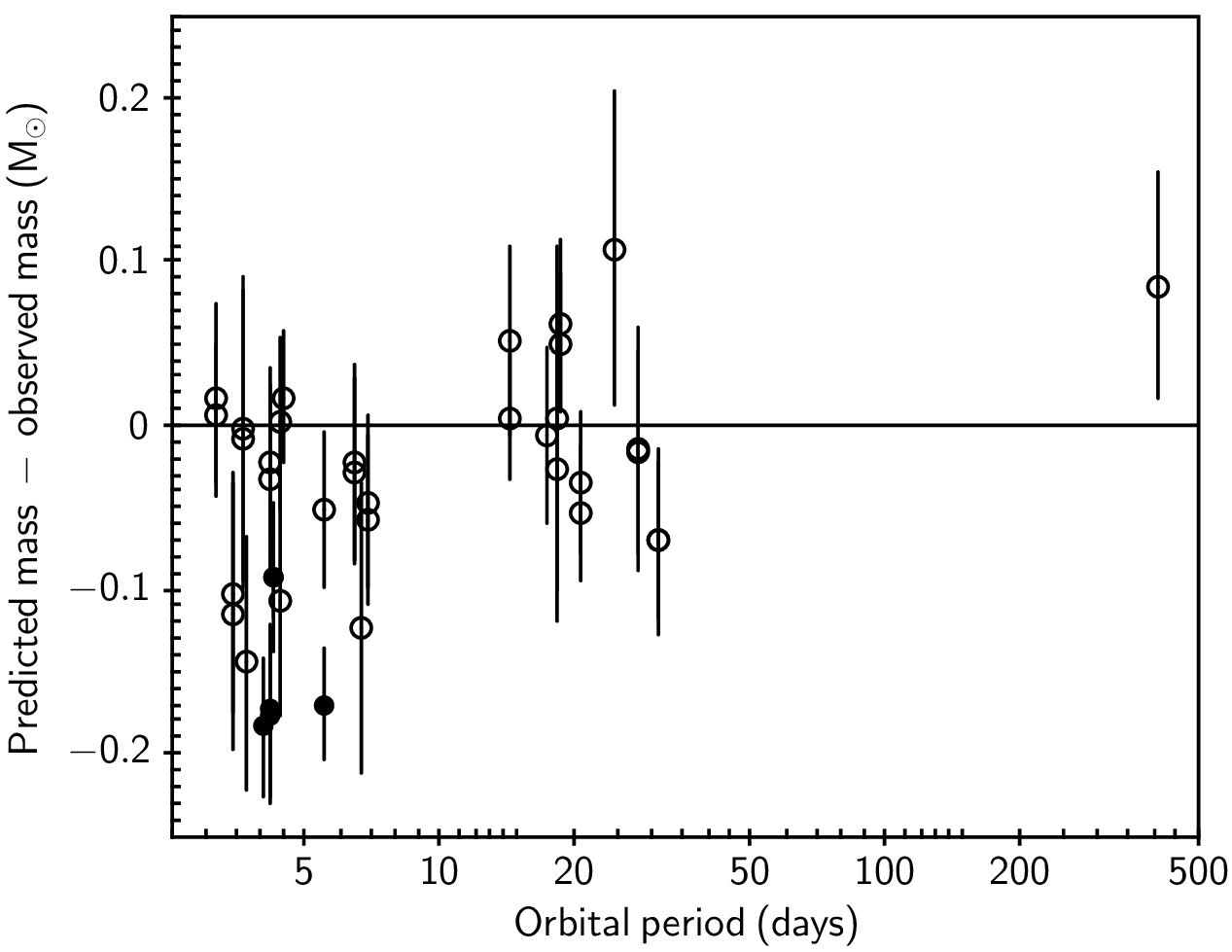}}
\caption{Error in the Bayesian mass estimate for stars in detached eclipsing
binary star systems as a function of orbital period. Errors in excess of 2
standard deviation are plotted with filled circles.
\label{debmass_pFig}}
\end{figure}

 We used DEBCat\footnote{\url{http://www.astro.keele.ac.uk/\~jkt/debcat/}} to
identify 39 stars in 24 detached eclipsing binary systems  that are suitable
for testing the accuracy of the mass estimated using our method when applied
in the mass range  typical for planet-host stars. The masses and radii of
the stars in this catalogue have been measured directly to a precision  of
better than 2\%. We have restricted our comparison to stars in binary systems
with orbital periods $P_\mathrm{orb} > 3$\,d in an attempt to avoid
complications that might arise due to the stars having strong tidal
interactions or very rapid rotation. We also exclude stars larger than
2\mbox{R$_{\sun}$}, stars without a measured value of
$\mathrm{[Fe/H]}_{\mathrm{s}}$ that include an estimate of the standard error,
and stars with $\mathrm{[Fe/H]}_{\mathrm{s}} < -0.4$. The limits on
R$_{\star}$  and $\mathrm{[Fe/H]}_{\mathrm{s}}$ were chosen so that the stars
are not close to the edge of the of the stellar model grid. For the majority
of planet host stars currently known the  luminosity of the star has not been
measured directly, so we do not include the luminosity of the eclipsing binary
stars as a constraint in this analysis. The properties of the stars in this
sample and the masses we derive using our method are given in
Table~\ref{DEBCatTable}. Kepler-34 and Kepler-35 are also planet host stars,
with planets on circumbinary orbits. 

 The observed and predicted masses are shown in Fig.~\ref{debmassFig}. There
is a clear tendency for our grid of standard stellar models to underpredict
the mass of some  stars by about 0.15\mbox{M$_{\sun}$}. We imposed an
arbitrary limit on the orbital period of the binary systems we have used so we
have investigated whether the orbital period is a factor in this analysis. The
approximate orbital period of each star is indicated by the plotting
symbol/colour in Fig.~\ref{debmassFig}. It is clear that all the stars with
discrepant masses have short orbital periods ($\la 6$\,d), but that not all
stars with short orbital periods have discrepant masses. This is more directly
seen in Fig.~\ref{debmass_pFig}, where we plot the mass discrepancy as a
function of orbital period. All the stars with a mass underpredicted by more
than 2 standard deviations have orbital periods $P_\mathrm{orb} < 7 $\,d, and
there is a clear tendency for the mass to be underpredicted for other stars
with orbital periods in this range. For low mass stars with large convective
envelopes in short-period binaries such as these, the rotational period of the
star and the orbital period are expected to be equal (or nearly so for
eccentric binaries) due to strong tidal interactions between the stars. This
has been confirmed by direct observation of detached eclipsing binary stars,
including some of the stars in this sample \citep{2010A+ARv..18...67T}.  If
we assume that rotation is the reason why the predicted masses of some stars
are discrepant, then we can conclude that our method is able to accurately
predict the mass within the stated uncertainties for stars with rotation
periods $P_\mathrm{rot} \ga 15 $\,d and in the mass range
0.8\,--\,1.3\,\mbox{M$_{\sun}$}. There are no suitable data for stars in
long-period binaries to test whether this conclusion holds in the mass range
1.3\,--\,1.6\,\mbox{M$_{\sun}$}. There are also no suitable data  to test our
method  in the mass range 0.6\,--\,0.8\,\mbox{M$_{\sun}$}, or for stars with
masses 0.8\,--\,1.3\,\mbox{M$_{\sun}$}\ and rotation periods $7\,\mathrm{d} <
P_\mathrm{rot} < 14$\,d.

 The binary star WOCS~40007 (\object{2MASS J19413393+4013003}) is a member of
the star cluster NGC~6819. \citet{2013AJ....146...58J} find the
age of NGC~6819 to be about 2.4\,Gyr from color-magnitude diagram
isochrone fitting and an age estimate for the binary system of $3.1 \pm
0.4$~Gyr based on stellar model fits to their mass and radius measurements of
the two stars. We obtain ages of $3.1 \pm 0.6$~Gyr and $3.3 \pm 1.8$~Gyr for
WOCS~40007~A and WOCS~40007~B, consistent with both the age
estimate for the binary system and the cluster  obtained by
\citeauthor{2013AJ....146...58J}

\begin{table*}
 \caption{Bayesian mass and age estimates for stars in detached eclipsing
binary star systems ($\langle \mbox{M$_{\star}$} \rangle$) compared to the mass of the
stars directly measured from observations ($M_\mathrm{obs}$). }
\label{DEBCatTable}
\centering
 \begin{tabular}{@{}lrrrrrrr}
\hline
\hline
Star &
  \multicolumn{1}{c}{P$_\mathrm{orb}$ } &
  \multicolumn{1}{c}{\mbox{T$_\mathrm{eff}$}} &
  \multicolumn{1}{c}{$\mathrm{[Fe/H]}_{\mathrm{s}}$}  &
  \multicolumn{1}{c}{\mbox{$\rho_{\star}$}} &
  \multicolumn{1}{c}{$M_\mathrm{obs}$} &
  \multicolumn{1}{c}{$\langle \mbox{M$_{\star}$} \rangle$ } & \\
 &
  \multicolumn{1}{c}{[d]} &
  \multicolumn{1}{c}{[K]} &
&
  \multicolumn{1}{c}{[$\rho_{\sun}$]} &
  \multicolumn{1}{c}{[\mbox{M$_{\sun}$}]} &
  \multicolumn{1}{c}{[\mbox{M$_{\sun}$}]} &
  \multicolumn{1}{l}{Ref.} \\
\hline
 \noalign{\smallskip}
WOCS 40007 A              & 3.18  & $6240 \pm 90  $&$   0.09 \pm  0.03 $&$  0.451\pm 0.010$&$ 1.236\pm 0.020 $&$ 1.243   \pm  0.038 $ & 1\\
WOCS 40007 B              &       & $5960 \pm 150 $&$                  $&$  0.820\pm 0.016$&$ 1.086\pm 0.018 $&$ 1.102   \pm  0.056 $ & 1 \\
 \noalign{\smallskip}
CD Tau A          & 3.43  & $6190 \pm 60  $&$   0.08 \pm  0.15 $&$  0.248\pm 0.008$&$ 1.442\pm 0.016 $&$ 1.326   \pm  0.079   $ & 2 \\
CD Tau B          &       & $6190 \pm 60  $&$                  $&$  0.344\pm 0.014$&$ 1.368\pm 0.016 $&$ 1.266   \pm  0.072   $ & 2 \\
 \noalign{\smallskip}
CO And A          & 3.65  & $6140 \pm 130 $&$   0.01 \pm  0.15 $&$  0.250\pm 0.009$&$ 1.289\pm 0.007 $&$ 1.281   \pm  0.090   $ & 3 \\
CO And  B          &       & $6160 \pm 130 $&$                  $&$  0.260\pm 0.008$&$ 1.264\pm 0.007 $&$ 1.263   \pm  0.092   $ & 3\\
 \noalign{\smallskip}
CoRoT 105906206 B & 3.69  & $6150 \pm 160 $&$   0.0  \pm  0.1  $&$ 0.536\pm 0.017$&$ 1.29 \pm 0.03  $&$ 1.145   \pm  0.071   $ & 4 \\
 \noalign{\smallskip}
GX Gem B          & 4.04  & $6160 \pm 100 $&$   -0.12\pm  0.10 $&$  0.131\pm 0.002$&$ 1.467\pm 0.010 $&$ 1.283   \pm  0.041   $ & 5\\
 \noalign{\smallskip}
UX Men A          & 4.18  & $6190 \pm 100 $&$   0.04 \pm  0.10 $&$  0.500\pm 0.015$&$ 1.223\pm 0.001 $&$ 1.191   \pm  0.056   $ &6,7 \\
        UX Men B          &       & $6150 \pm 100 $&$                  $&$  0.574\pm 0.018$&$ 1.188\pm 0.001 $&$ 1.165   \pm  0.057   $ &6,7 \\
 \noalign{\smallskip}
WZ Oph A          & 4.18  & $6160 \pm 100 $&$   -0.27\pm  0.07 $&$  0.446\pm 0.012$&$ 1.227\pm 0.007 $&$ 1.054   \pm  0.052   $ & 8,9 \\
WZ Oph B          &       & $6110 \pm 100 $&$                  $&$  0.427\pm 0.011$&$ 1.22 \pm 0.006 $&$ 1.043   \pm  0.053   $ & 8,9 \\
 \noalign{\smallskip}
V636 Cen A        & 4.28  & $5900 \pm 80  $&$   -0.20\pm  0.08 $&$  0.997\pm 0.013$&$ 1.052\pm 0.005 $&$ 0.960   \pm  0.045   $ & 10,11 \\
V636 Cen B        &       & $5000 \pm 100 $&$                  $&$  1.493\pm 0.022$&$ 0.854\pm 0.003 $&$ 0.762   \pm  0.024   $ & 10,11 \\
 \noalign{\smallskip}
CoRoT 102918586 B & 4.39  & $7100 \pm 120 $&$   0.11 \pm  0.05 $&$ 0.460\pm 0.013$&$ 1.49 \pm 0.03  $&$ 1.492   \pm  0.043   $ & 12 \\
 \noalign{\smallskip}
EK Cep B          & 4.43  & $5700 \pm 200 $&$   0.07 \pm  0.05 $&$  0.494\pm 0.009$&$ 1.124\pm 0.012 $&$ 1.017   \pm  0.069   $ & 13 \\
 \noalign{\smallskip}
YZ Cas B          & 4.47  & $6890 \pm 240 $&$   0.10 \pm  0.06 $&$  0.562\pm 0.008$&$ 1.325\pm 0.007 $&$ 1.342   \pm  0.040   $ & 14 \\
 \noalign{\smallskip}
BK Peg A          & 5.49  & $6270 \pm 90  $&$   -0.12\pm  0.07 $&$  0.180\pm 0.002$&$ 1.414\pm 0.007 $&$ 1.244   \pm  0.033   $ & 15 \\ 
BK Peg B          &       & $6320 \pm 90  $&$                  $&$  0.392\pm 0.013$&$ 1.257\pm 0.005 $&$ 1.205   \pm  0.047   $ & 15 \\
 \noalign{\smallskip}
V785 Cep A        & 6.50  & $5900 \pm 100 $&$   -0.06\pm  0.06 $&$  0.382\pm 0.015$&$ 1.103\pm 0.007 $&$ 1.081   \pm  0.059   $ & 16 \\ 
V785 Cep B        &       & $5870 \pm 100 $&$                  $&$  0.418\pm 0.017$&$ 1.081\pm 0.007 $&$ 1.053   \pm  0.057   $ & 16 \\
 \noalign{\smallskip}
BW Aqr B          & 6.72  & $6220 \pm 100 $&$   -0.07\pm  0.11 $&$  0.242\pm 0.018$&$ 1.386\pm 0.021 $&$ 1.262   \pm  0.086   $ & 17 \\ 
 \noalign{\smallskip}
EW Ori A          & 6.94  & $6070 \pm 100 $&$   0.05 \pm  0.09 $&$  0.736\pm 0.012$&$ 1.173\pm 0.011 $&$ 1.126   \pm  0.051   $  & 18 \\ 
EW Ori B          &       & $5900 \pm 100 $&$                  $&$  0.851\pm 0.014$&$ 1.123\pm 0.009 $&$ 1.066   \pm  0.051   $  & 18 \\
 \noalign{\smallskip}
V568 Lyr A        & 14.47 & $5650 \pm 90  $&$   0.28 \pm  0.05 $&$  0.399\pm 0.011$&$ 1.087\pm 0.004 $&$ 1.138   \pm  0.058   $  & 19 \\ 
V568 Lyr B        &       & $4820 \pm 150 $&$                  $&$  1.735\pm 0.034$&$ 0.828\pm 0.002 $&$ 0.832   \pm  0.038   $  & 19 \\
 \noalign{\smallskip}
KIC 6131659 A     &17.53& $5790 \pm 50  $&$   -0.23\pm  0.20 $&$  1.353\pm 0.016$&$ 0.922\pm 0.007 $&$ 0.915   \pm  0.053   $  & 20 \\ 
 \noalign{\smallskip}
LV Her A          & 18.44 & $6210 \pm 160 $&$   0.08 \pm  0.21 $&$  0.476\pm 0.013$&$ 1.193\pm 0.010 $&$ 1.197   \pm  0.104   $ & 21\\ 
LV Her B          &       & $6020 \pm 160 $&$                  $&$  0.517\pm 0.013$&$ 1.170\pm 0.008 $&$ 1.142   \pm  0.092   $ & 21 \\
 \noalign{\smallskip}
V565 Lyr A        & 18.80 & $5600 \pm 90  $&$   0.28 \pm  0.05 $&$  0.746\pm 0.014$&$ 0.995\pm 0.003 $&$ 1.045   \pm  0.042   $ & 22 \\ 
V565 Lyr B        &       & $5430 \pm 130 $&$                  $&$  1.016\pm 0.017$&$ 0.929\pm 0.003 $&$ 0.991   \pm  0.051   $ & 22 \\
 \noalign{\smallskip}
Kepler-35 A       & 20.73 & $5610 \pm 140 $&$   -0.34\pm  0.20 $&$  0.816\pm 0.007$&$ 0.888\pm 0.005 $&$ 0.835   \pm  0.041   $  & 23 \\
Kepler-35 B       &       & $5200 \pm 100 $&$                  $&$  1.666\pm 0.016$&$ 0.809\pm 0.005 $&$ 0.775   \pm  0.043   $  & 23 \\
 \noalign{\smallskip}
AI Phe B          & 24.59 & $6310 \pm 150 $&$   -0.14\pm  0.10 $&$  0.200\pm 0.008$&$ 1.195\pm 0.004 $&$ 1.303   \pm  0.096   $  & 24 \\ 
 \noalign{\smallskip}
Kepler-34 A       & 27.80 & $5920 \pm 120 $&$   -0.07\pm  0.15 $&$  0.668\pm 0.006$&$ 1.048\pm 0.003 $&$ 1.033   \pm  0.074   $  & 25 \\ 
Kepler-34 B       &       & $5860 \pm 140 $&$                  $&$  0.782\pm 0.007$&$ 1.021\pm 0.002 $&$ 1.004   \pm  0.061   $  & 25 \\
 \noalign{\smallskip}
KX Cnc A          & 31.22 & $5900 \pm 100 $&$   0.07 \pm  0.10 $&$  0.945\pm 0.006$&$ 1.138\pm 0.003 $&$ 1.067   \pm  0.057   $  & 26 \\ 
KX Cnc B          &       & $5850 \pm 100 $&$                  $&$  0.980\pm 0.006$&$ 1.131\pm 0.003 $&$ 1.061   \pm  0.048   $  & 26 \\
 \noalign{\smallskip}
KIC 8410637 B     & 408.3 & $6490 \pm 170 $&$   0.20 \pm  0.11 $&$ 0.341\pm 0.021$&$ 1.322\pm 0.017 $&$ 1.407   \pm  0.067   $  & 27 \\ 
 \noalign{\smallskip}
\hline
 \end{tabular}   
\tablebib{
(1)~\citet{2013AJ....146...58J}; (2)~\citet{1999MNRAS.309..199R}; (3)~\citet{2010AJ....139.2347L}; (4)~{\citet{2014A+A...565A..55D}};
(5)~\citet{2008AJ....135.1757L}; (6)~{\citet{1989A+A...211..346A}} (7)~\citet{2009MNRAS.400..969H};(8)~{\citet{2008A+A...487.1081C}};
(9)~{\citet{2008A+A...487.1095C}}; (10)~{\citet{2008A+A...487.1081C}}; (11)~{\citet{2009A+A...502..253C}}; (12)~{\citet{2013A+A...552A..60M}} 
(13)~{\citet{1993A+A...274..274M}}; (14)~\citet{2014MNRAS.438..590P}; (15)~{\citet{2010A+A...516A..42C}}; (16)~\citet{2009AJ....137.5086M};
(17)~{\citet{2010A+A...516A..42C}}; (18)~{\citet{2010A+A...511A..22C}}; (19)~{\citet{2011A+A...525A...2B}}  ; (20)~\citet{2012ApJ...761..157B};
(21)~\citet{2009AJ....138.1622T}; (22)~{\citet{2011A+A...525A...2B}} ; (23)~\citet{2012Natur.481..475W}; (24)~{\citet{1988A+A...196..128A}};
(25)~\citet{2012Natur.481..475W}; (26)~\citet{2012AJ....143....5S}; (27)~{\citet{2013A+A...556A.138F}}   
}
 \end{table*}

\subsection{Systematic errors - helium abundance and mixing length}

\begin{figure}
\resizebox{\hsize}{!}{\includegraphics{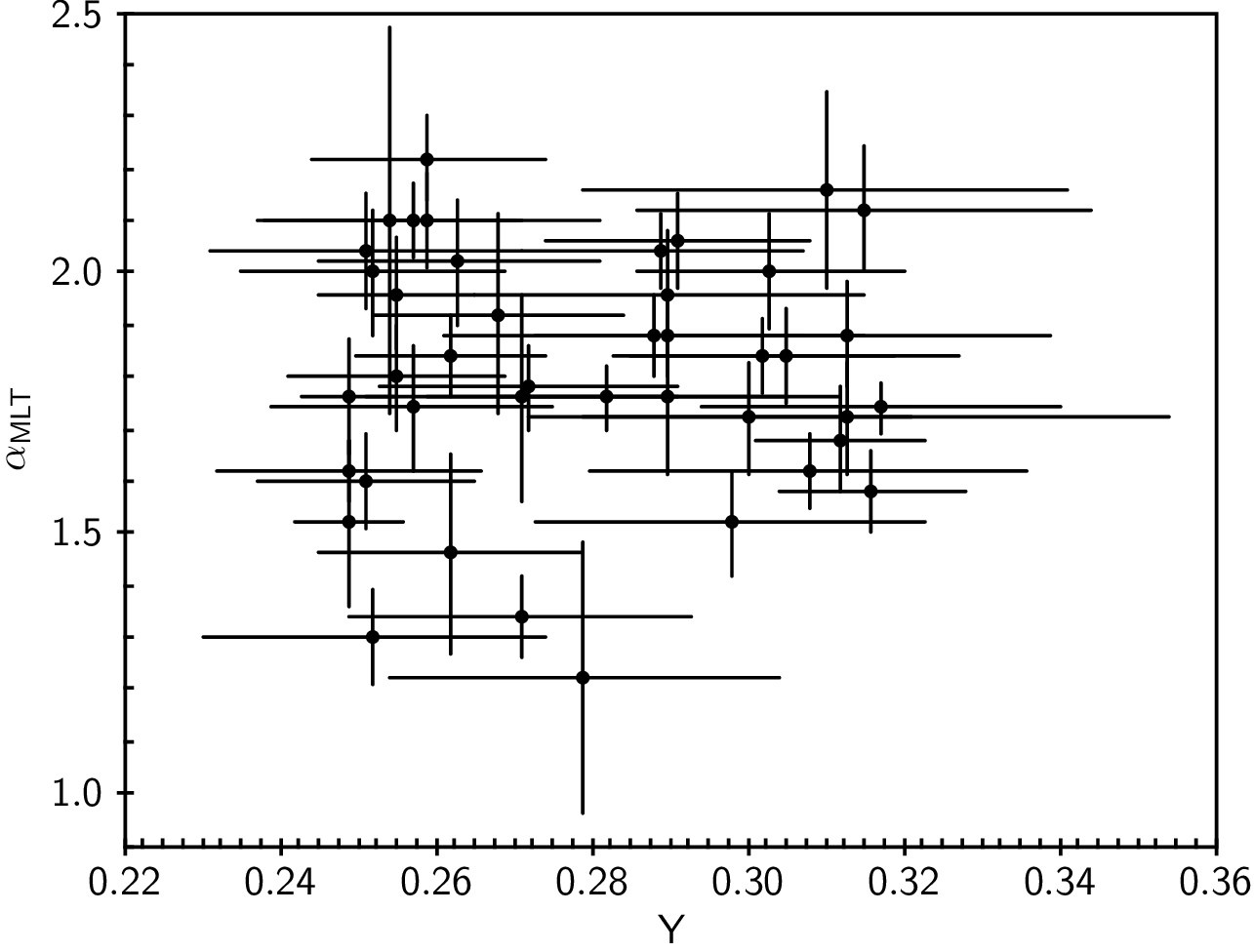}}
\caption{Helium abundance $Y$ and mixing length parameter $\alpha_\mathrm{MLT}$ for 42
solar-like stars from \protect{\citet{2014arXiv1402.3614M}}. 
\label{YamltFig}}
\end{figure}

 The results derived using our grid of stellar models with fixed values of
$\alpha_\mathrm{MLT}$ and $Y$ calibrated on the Sun are subject to some level
of systematic error due to the variations in these values between different
stars. Fortunately, observed constraints on the variation in both of these
factors  are now available thanks to Kepler light curves of sufficient quality
to enable the study of solar-like oscillations for a large sample of late-type
stars. The Kepler data for 42 solar-type stars have been analysed by
\citet{2014arXiv1402.3614M} including $\alpha_\mathrm{MLT}$ and $Y$ as free
parameters in the fit of the stellar models to observed frequency spectrum.
The values of $Y$ and $\alpha_\mathrm{MLT}$ derived from the asteroseismology
are shown in Fig.~\ref{YamltFig}. It can be seen that there is no strong
correlation between $Y$ and $\alpha_\mathrm{MLT}$, and that there is additional
scatter in these values beyond the quoted standard errors. If we add 0.02 in
quadrature to the standard errors on $Y$ then a least-squares fit of a
constant to the values of $Y$ gives a reduced chi-squared value $\chi^2_{\rm
r}\approx 1$, so we use 0.02 as an estimate of the scatter in $Y$. Similarly,
for $\alpha_\mathrm{MLT}$ we assume that the scatter around the solar-calibrated
value is 0.2. It is debatable whether these derived values of $Y$ are reliable
measurements of the actual helium abundance of these stars or whether the
derived values of $\alpha_\mathrm{MLT}$ are an accurate reflection of the
properties of convection in their atmospheres. However,  the frequency
spectrum of solar-type stars is very sensitive to the mean density of the star
so it is reasonable to use the observed scatter in $Y$ and $\alpha_\mathrm{MLT}$
from the study of \citeauthor{2014arXiv1402.3614M} as a way to estimate the
systematic error due to the uncertainties in these values in the masses and
ages derived using our method.

 In principle it may be possible to estimate an appropriate value of
$\alpha_\mathrm{MLT}$ to use for a given star based on an empirical or
theoretical calibration of $\alpha_\mathrm{MLT}$ against properties such as
mass, surface gravity, etc. We have decided not to attempt this and instead to
treat the  scatter in $\alpha_\mathrm{MLT}$ and $Y$ as a sources of possible
systematic error in the masses and ages. The systematic error in the mass due
to the uncertainty on $Y$ is given by the quantity $\sigma_{M, Y}$, which is
the change in in the best-fit mass, $M_\mathrm{b}$, produced by applying our
method using the grid of stellar models in which the helium abundance is
increased by $+0.02$ compared to the value used in our grid of standard
stellar models.  Similarly, $\sigma_{\tau, Y}$ is the systematic error in the
age due to the uncertainty on $Y$ calculated in the same way. For the
systematic errors in the mass and age due to the uncertainty in
$\alpha_\mathrm{MLT}$ we calculate the best-fitting mass and age for a grid of
stellar models calculated with {\sc garstec} for $\alpha_\mathrm{MLT} = 1.50$
and multiply the resulting change in mass or age by the factor
$0.2/(1.50-1.78)$, i.e, $\sigma_{M,\mathrm{\alpha}}$ is the change in the
best-fitting mass due to an increase of 0.2 in $\alpha_\mathrm{MLT}$ and
similarly for the age and $\sigma_{\tau,\mathrm{\alpha}}$. The two grids of
stellar models necessary for these calculations are included in the version of
{\sc bagemass} that accompanies this article so the user is free to decide
whether they should account for this potential source of additional
uncertainty for the star they are analysing.

\section{Results}

 We have applied our method to over 200 stars that host transiting planet or
brown dwarf companions and have found that the software runs without problems
in all these cases and that our results are generally in good agreement with
published results. Here we only report the results for a selection of stars to
illustrate the main features of our method and to highlight some interesting
results.

The input data used for our analysis is given in Table~\ref{DataTable}. For
stars that have a trigonometrical parallax in \citet{2007A+A...474..653V}
with precision $\sigma_{\pi}/\pi \la 0.1$ we have calculated L$_{\star}$ using
a value of $f_{\oplus}$ estimated by integrating a synthetic stellar spectrum
fit by least-squares to the observed fluxes of the star.  Optical photometry
is obtained from the Naval Observatory Merged Astrometric Dataset (NOMAD)
catalogue\footnote{\url{http://www.nofs.navy.mil/data/fchpix}}
\citep{2004AAS...205.4815Z}, the Tycho-2 catalogue
\citep{2000A+A...355L..27H} and the Carlsberg Meridian Catalog 14
\citep{2006yCat.1304....0C}. Near-infrared photometry was obtained from the
Two Micron All Sky Survey
(2MASS)\footnote{\url{http://www.ipac.caltech.edu/2mass}} and Deep Near
Infrared Survey of the Southern Sky
(DENIS)\footnote{\url{http://cdsweb.u-strasbg.fr/denis.html}} catalogues
\citep{2005yCat.2263....0T,2006AJ....131.1163S}. The synthetic stellar
spectra used for the numerical integration of the fluxes are from
\citet{1993KurCD..13.....K}. Reddening can be neglected for these nearby
stars given the accuracy of the measured fluxes and parallaxes.  Standard
errors are estimated using a simple Monte Carlo method in which we generate
65,536 pairs of $\pi$ and $f_{\oplus}$ values from Gaussian distributions and
then find the 68.3\% confidence interval of the resulting $\log(L/L_{\sun})$
values. 

 The results of the analysis for our selection of planet host stars  are given
in Table~\ref{ResultsTable}, where we provide the values of the mass, age and
initial metallicity that provide the best fit to the observed properties of
the star, $M_\mathrm{b}$, $\tau_\mathrm{b}$ and
$\mathrm{[Fe/H]}_{\mathrm{b}}$, respectively, the value of $\chi^2$ for this
solution, and the mean and standard deviation for each of the posterior
distributions of  age and mass. The likely evolutionary state of the star is
indicated by the quantity $p_{\rm MS}$, which is the fraction of points in the
chain for which the central hydrogen abundance is not 0, i.e., $p_\mathrm{MS}$
is the probability that the star is still on the main-sequence.

\begin{figure*}
\resizebox{\hsize}{!}{\includegraphics{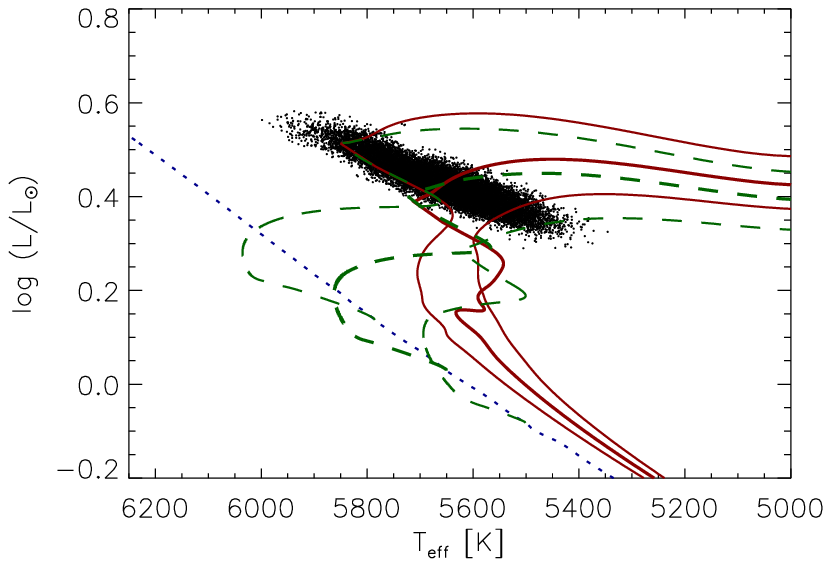}
\includegraphics{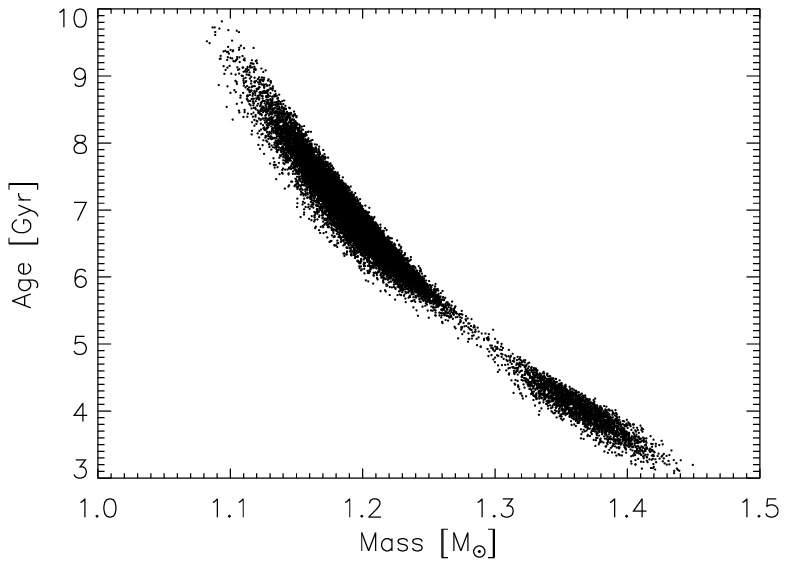}}
\caption{Left panel: Results of our MCMC analysis for HAT-P-13 in the
Hertzsprung-Russell diagram. Black dots are individual steps in the Markov
chain. The dotted line (blue) is the ZAMS. Solid
lines (red) are isochrones for ages $6.6\pm1.2$\,Gyr.  Dashed lines (green)
are evolutionary tracks for masses $1.21\pm0.06$\,\mbox{M$_{\sun}$}. All isochrones and
tracks are for $\mathrm{[Fe/H]}_{\mathrm{i}}=0.47$. Right panel: Joint posterior
distribution for the mass and age and HAT-P-13 from our MCMC analysis.  
\label{hatp13massage}}
\end{figure*}
 
\begin{table*}
\caption{Observed properties of a selection of stars that host transiting
extrasolar planets. }
\label{DataTable}
\centering
 \begin{tabular}{@{}lrrrrrrr}
\hline
\hline
Star & \multicolumn{1}{l}{P [d]}&\multicolumn{1}{l}{\mbox{T$_\mathrm{eff}$} [K]}&\multicolumn{1}{l}{[Fe/H]}
&\multicolumn{1}{l}{$\rho/\rho_{\sun}$} & \multicolumn{1}{l}{$f_{\oplus}$ [pW\,m$^{-2}$]} &
\multicolumn{1}{l}{$\log(L_{\star}/L_{\sun})$} & Ref.\\
\noalign{\smallskip}
\hline
HAT-P-13 = TYC 3416-543-1  & 2.92 &$ 5653 \pm 90 $&$ +0.41 \pm 0.08 $&$ 0.244 \pm 0.013  $& & & 1 \\
HD 209458 & 3.52 &$ 6117 \pm 50 $&$ +0.02 \pm 0.05 $&$ 0.733 \pm 0.008  $& $ 23.1 \pm 1.2   $&$ +0.25\pm 0.04  $ & 2\\
WASP-32   & 2.72 &$ 6042 \pm 47 $&$ -0.07 \pm 0.09 $&$ 0.840 \pm 0.050  $& & & 3,4 \\
HD 189733 & 2.22 &$ 5050 \pm 50 $&$ -0.03 \pm 0.05 $&$ 1.980 \pm 0.170  $& $ 27.5 \pm 1.4   $&$ -0.49\pm 0.025 $ & 2 \\
WASP-52   & 1.75 &$ 5000 \pm 100 $&$ +0.03 \pm 0.12 $&$ 1.760 \pm 0.080  $&&&5\\
Qatar 2   & 1.34 &$ 4645 \pm 50 $&$ -0.02 \pm 0.08 $&$ 1.591 \pm 0.016  $& & & 6\\
\hline
\end{tabular}   
\tablebib{
(1)~\citet{2012MNRAS.420.2580S};
(2)~\citet{2010MNRAS.408.1689S};
(3)~\citet{2012ApJ...760..139B};
(4)~\citet{2014ApJ...788...39T};
(5)~{\citet{2013A+A...549A.134H}};
(6)~\citet{2014arXiv1406.6714M}.
}
\end{table*}     

\begin{table*}
 \caption{Bayesian mass and age estimates for a selection host stars of
transiting extrasolar planets.}
\label{ResultsTable}
\centering
\begin{tabular}{@{}lrrrrrrrrrrrr}
\hline
\hline
Star &
  \multicolumn{1}{c}{$\tau_\mathrm{b}$ [Gyr]} &
  \multicolumn{1}{c}{$M_\mathrm{b}$[M$_{\sun}$]} &
  \multicolumn{1}{c}{$\mathrm{[Fe/H]}_\mathrm{i, b}$} &
  \multicolumn{1}{c}{$\chi^2$}&
  \multicolumn{1}{c}{$\langle \tau_{\star} \rangle$ [Gyr]}  &
  \multicolumn{1}{c}{$\langle \mathrm{M}_{\star} \rangle$ [M$_{\sun}$]} &
  \multicolumn{1}{c}{$p_\mathrm{MS}$}& 
  \multicolumn{1}{c}{$\sigma_{\tau, Y}$}  &
  \multicolumn{1}{c}{$\sigma_{\tau,\alpha}$} &
  \multicolumn{1}{c}{$\sigma_{M, Y}$}  &
  \multicolumn{1}{c}{$\sigma_{M,\alpha}$}  \\
\hline
 \noalign{\smallskip}
HAT-P-13  &  6.59 &  1.206 &$ +0.473 $&  0.02 &$ 6.3 \pm  1.4 $&$  1.23 \pm 0.08 $& 0.29 &$ -0.05 $&$  1.74 $&$ -0.042 $&$ -0.087 $ \\
HD 209458 &  2.39 &  1.142 &$ +0.068 $&  0.19 &$ 2.4 \pm  0.8 $&$  1.14 \pm 0.04 $& 1.00 &$  0.20 $&$  1.29 $&$ -0.038 $&$ -0.033 $ \\
WASP-32   &  3.35 &  1.062 &$ -0.025 $&  0.01 &$ 3.5 \pm  1.4 $&$  1.10 \pm 0.05 $& 1.00 &$  0.47 $&$  1.64 $&$ -0.045 $&$ -0.032 $ \\
HD 189733 &  1.58 &  0.824 &$ -0.008 $&  0.02 &$ 4.9 \pm  3.3 $&$  0.80 \pm 0.02 $& 1.00 &$  2.32 $&$  1.12 $&$ -0.036 $&$ -0.007 $ \\
WASP-52   &  5.99 &  0.820 &$ +0.078 $&  0.01 &$ 7.1 \pm  3.6 $&$  0.81 \pm 0.04 $& 1.00 &$  2.53 $&$  3.46 $&$ -0.043 $&$ -0.022 $ \\
Qatar 2   & 17.50 &  0.751 &$ +0.149 $&  3.58 &$15.8 \pm  1.4 $&$  0.77 \pm 0.01 $& 1.00 &$ -0.04 $&$  0.04 $&$ -0.018 $&$  0.013 $ \\
\hline
 \end{tabular}   
 \end{table*}     

\section{Discussion}

 An example of applying {\sc bagemass} to one star (\object{HAT-P-13}) is
shown in Fig.~\ref{hatp13massage}. We have chosen this star as an example
because it shows the difficulties that can arise in the analysis of stars
close to the end of their main sequence evolution. The complexity of the
evolution tracks and isochrones in the Hertzsprung-Russell diagram is clear,
even though we have restricted the models plotted to one value of
$\mathrm{[Fe/H]}_{\mathrm{i}}$ (the best-fit value).  The joint posterior
distribution for the mass and age is clearly bimodal, with peaks in the
distribution near ($M, \tau_{\star})  = (1.20\,\mbox{M$_{\sun}$}, 6.7$~Gyr)
and (1.37\,\mbox{M$_{\sun}$}, 4.1\,Gyr). It can be difficult to ensure that
the Markov chain has converged for distributions of this type where there are
two solutions separated in the parameter space by a region of low likelihood.
We used another MCMC analysis for this star with 500,000 steps in the chain to
verify that the default chain length of 50,000 is sufficient even in this
difficult case to produce an accurate estimate of the posterior probability
distribution. The maximum-likelihood solution in this case has a mass  and age
$(M_\mathrm{b},\tau_\mathrm{b})= (1.21\,\mbox{M$_{\sun}$},
6.5\,\mathrm{Gyr})$. The mean and standard deviation of the joint posterior
distribution for the mass and age are $(\langle \mbox{M$_{\star}$}
\rangle,\langle \tau_{\star} \rangle) = (1.21\pm0.06\,\mbox{M$_{\sun}$},
6.7\pm1.2\,\mathrm{Gyr})$. The good agreement between the maximum-likelihood
solution and Bayesian analysis is a consequence of the low uncertainty in the
value of \mbox{$\rho_{\star}$}. In logarithmic terms the standard error on
\mbox{$\rho_{\star}$}\ for this star is 0.023\,dex. This is much lower than
the typical uncertainty  for estimates of $\log g$ based on the spectroscopic
analysis of a late-type star \citep[$\approx
0.1$\,dex,][]{2013MNRAS.428.3164D}. In general, we should expect that the
terminal age bias for planet host stars for isochrone fitting using the
density is much less severe than for single stars using $\log g$. Even so, the
full Bayesian analysis is worthwhile to obtain an objective estimate of the
true range of possible masses and ages for a planet host star, particularly
for stars near the main sequence turn off or if the error on
\mbox{$\rho_{\star}$}\ is large. We have also found that the joint probability
distribution for mass and age from the full Bayesian analysis can also be
useful for improving the power of statistical tests based on these quantities,
e.g., comparing the ages derived from stellar models to  gyrochronological
ages (Maxted et al., in prep.).

\object{HD 209458} was the first transiting exoplanet discovered
\citep{2000ApJ...529L..45C} and is also one of the brightest and
best-studied. The properties of this planetary system are also typical
for hot Jupiter exoplanets. The fit to the observed properties of this star
including the luminosity constraint is good ($\chi^2 = 0.19$ for one degree of
freedom). It is interesting to compare the results for HD~209458 to
those for \object{WASP-32}, which is also a typical hot Jupiter system but one
for which there is currently no accurate parallax measurement. If we compare
the values of $\langle \tau_{\star} \rangle$ and $\langle \mathrm{M}_{\star}
\rangle$ for these stars in Table~\ref{ResultsTable} we see that the adding
parallax measurement with the best precision currently available does not lead
to a significant improvement in the precision of the mass and age estimates --
the slightly better precision of the age estimate for HD~209458 is
mostly due to the higher precision of the density estimate. A similar argument
applies in the case of the stars \object{HD 189733} and \object{WASP-52}.
These stars are less massive than HD~209458 and close to the limit at
which age estimates from stellar models become impossible because there is no
significant evolution of the star during the lifetime of the Galaxy.

 For stars without a parallax measurement the number of observables
is the same as the number of model parameters, but it is still possible that
no models in the standard model grid give a good fit to the observed
properties of the star. This is the case for \object{Qatar 2}, which is a good
example of a star that appears to be older than the Galactic disc
\citep[10\,Gyr,][]{2014A+A...566A..81C}. We used the Markov chain
calculated for this star to estimate an upper limit to the probability that
the age of this star derived from our stellar models is less than 10\,Gyr and
find $P(\tau_{\star} < 10\,{\rm Gyr}) < 0.002$. This probability is an upper
limit because the best-fit to the observed parameters occurs at the edge of
the model grid ($\tau_\mathrm{b} = 17.5$\,Gyr). This  also means that the
values of $\sigma_{\tau, Y}$ and $\sigma_{\tau,{\rm \alpha}}$ are not reliable
in this case. 

 It has long been known that some K-dwarfs appear to be larger by 5\% or more
than the radius predicted by standard stellar models
\citep{1973A+A....26..437H, 1997AJ....114.1195P}. This  ``radius anomaly''
is correlated with the rotation rate of the star, but also shows some
dependence on the mass and metallicity of the star \citep{2007ApJ...660..732L,
2013ApJ...776...87S}. The dependence on rotation is thought to be the result
of the increase in magnetic activity for rapidly rotating stars. Magnetic
activity can affect the structure of a star by producing a high coverage of
starspots, which changes the  boundary conditions at the surface of the star,
or by reducing the efficiency of energy transport by convection. Whatever the
cause of the radius anomaly in K-dwarfs, the existence of inflated K-dwarfs is
likely to be the reason why Qatar~2 has an apparent age  $> 10$\,Gyr. One
method that has been proposed to deal with the radius anomaly is to simulate
the magnetic inhibition of convection by reducing the mixing length parameter
\citep{2007A+A...472L..17C}. For Qatar~2, we found that models with
$\alpha_\mathrm{MLT} < 1.4$ can match the properties of this star for ages
less than 10\,Gyr. This phenomenological approach has some support from
stellar models that incorporate magnetic fields in a self-consistent way
\citep{2013ApJ...779..183F}.  There is currently no well-established method to
predict the correct value of $\alpha_\mathrm{MLT}$ to use for a given star
affected by the radius anomaly and so no way to estimate the ages of such
stars.

\citet{2010A+A...516A..42C} found a reasonable match to the observed masses
and radii of both stars in the eclipsing binary system \object{BK Peg} using
both the VRSS stellar evolution models \citep{2006ApJS..162..375V} and the
Yale-Yonsei ``$Y^2$'' models \citep{ 2004ApJS..155..667D}, so the difference
between the observed mass of BK~Peg~A and the predicted mass with our grid of
standard stellar models needs some explanation. Both model grids used by
\citet{2010A+A...516A..42C} use old
estimates for the reaction rate  $^{14}{\rm N} ( {\rm p},\gamma) ^{15}{\rm O}
$. This reaction is the bottle-neck in the CNO cycle and the relevant reaction
rate has been redetermined experimentally and theoretically in the last
decade, resulting in a reduction by a factor of 2 compared to previous values
\citep{2011RvMP...83..195A}. We are able to reproduce this good fit to the
mass and radius of BK~Peg~A with {\sc garstec} if we use the old (inaccurate)
reaction rate. With the new reaction rate and the correct value of the star's
mass  the ``red-hook'' in the stellar evolution track as the star leaves the
main sequence does not reach values of T$_\mathrm{eff}$ low enough to match
the observed T$_\mathrm{eff}$ value. We also experimented with variations in
the assumed convective overshooting parameter, $\alpha_\mathrm{ov}$, but this
does not help to produce a good fit.

Decreasing $Y$ or, to a lesser extent, $\alpha_\mathrm{MLT}$ leads to an
increase in the estimated mass for stars with mass $\goa 0.9 M_{\sun}$. This
suggests that a comprehensive analysis similar to the one presented here but
applied to eclipsing binary stars will yield useful insights into stellar
models for such stars. Indeed, \citet{2012MNRAS.425.3104F} have undertaken
just such an analysis using a maximum likelihood method to estimate the helium
abundance and mixing length parameter for 14 stars in seven eclipsing binary
stars with masses from 0.85\,$M_{\sun}$ to 1.27\,$M_{\sun}$.  They find a weak
trend for stars with large rotation velocities to be fit better by models with
lower values of $\alpha_\mathrm{MLT}$, in qualitative agreement with the
results that we have found here. We have assumed that rotation is also the
reason why the mass esimated using our method is too low by about
0.1\,M$_{\sun}$ for some stars with masses $\ga 1 M_{\sun}$ in detached
eclipsing binaries. This issue deserves further investigation, ideally using
Bayesian methods similar to those developed here. Given the number of model
parameters and other factors such as rotation and magnetic activity that
should be considered, there is a clear need for accurate mass, radius,
T$_\mathrm{eff}$ and [Fe/H] measurements for more eclipsing binary stars. In
particular, there is a lack of data for stars in long period eclipsing
binaries  with masses 1.3\,--\,1.6\,M$_{\sun}$, i.e, at the upper end of the
mass distribution for planet host stars. 

\section{Conclusion}
 The software used to produce the results in this paper is written in standard
{\sc fortran}, is freely available and the installation requires only one,
widely-used additional software library \citep[{\sc fitsio},
][]{1998ASPC..145...97P}. The method has been validated against other models
and tested using the observed properties of detached eclipsing binary stars.
The method and its limitations are fully described herein. As a result, it is
now straightforward to determine accurately the joint posterior probability
distribution for the mass and age of a star eclipsed by a planet or other
dark body based on its observed properties and a state-of-the art set of
stellar models.

\begin{acknowledgements}
 JS acknowledges financial support from the Science and Technology Facilities
Council (STFC) in the form of an Advanced Fellowship. AMS is supported by the
MICINN ESP2013-41268-R grant and the Generalitat of Catalunya program
SGR-1458.

 The {\sc pspline} package is provided by the National Transport Code
Collaboration (NTCC),  a U.S. Department of Energy (DOE) supported activity
designed to facilitate research towards development of fusion energy as a
potential energy source for the future.  The NTCC website is supported by
personnel at the Princeton Plasma Physics Laboratory (PPPL), under grants and
contracts funded by the US Department of Energy. 
\end{acknowledgements}

\bibliographystyle{aa} 
\bibliography{wasp}
\end{document}